\documentclass[useAMS,usenatbib]{mn2e}
\usepackage{color}
\usepackage{graphicx}
\usepackage{xspace}
\usepackage{url}
\usepackage{natbib}
\usepackage{amsmath}

\title[\Ktwo systematics correction]{K2SC: Flexible systematics correction and detrending of \Ktwo light curves using Gaussian Process regression}
\author[S. Aigrain et al.]{
S. Aigrain\thanks{E-mail: suzanne.aigrain@astro.ox.ac.uk}, H. Parviainen and B. J. S. Pope\\
Department of Physics, University of Oxford, Keble Road, Oxford OX3 9UU, UK}

\newcommand{\cmat}{\ensuremath{\mathbf{K}}\xspace}
\newcommand{\flux}{\ensuremath{\mathbf{f}}\xspace}
\newcommand{\Ktwo}{\textit{K2}\xspace}
\newcommand{\ksc}{\textsc{k2sc}\xspace}
\newcommand{\kvc}{\textsc{k2varcat}\xspace}
\newcommand{\ksf}{\textsc{k2sff}\xspace}
\newcommand{\sap}{\textsc{sap}\xspace}
\newcommand{\pdc}{\textsc{pdc}\xspace}

\begin{document}
\date{Accepted \ldots Received \ldots; in original form \ldots}
\pagerange{\pageref{firstpage}--\pageref{lastpage}} \pubyear{2016}

\maketitle

\label{firstpage}

\begin{abstract}
We present \ksc (\Ktwo Systematics Correction), a {\sc Python} pipeline to model instrumental systematics and astrophysical variability in light curves from the \Ktwo mission. \ksc uses Gaussian process regression to model position-dependent systematics and time-dependent variability simultaneously, enabling the user to remove both (e.g., for transit searches) or to remove systematics while preserving variability (for variability studies). For  periodic variables, \ksc automatically computes estimates of the period, amplitude and evolution timescale of the variability. We apply \ksc to publicly available \Ktwo data from campaigns 3--5 showing that we obtain photometric precision approaching that of the original \emph{Kepler} mission. We compare our results to other publicly available \Ktwo pipelines, showing that we obtain similar or better results, on average. We use transit injection and recovery tests to evaluate the impact of \ksc on planetary transit searches in \Ktwo \pdc (Pre-search Data Conditioning) data, for 
planet-to-star radius ratio down $R_{\rm p}/R_\star=0.01$ and periods up to $P=40$\,d, and show that \ksc significantly improves the ability to distinguish between correct and false detections, particularly for small planets.
\ksc can be run automatically on many light curves, or manually tailored for specific objects such as pulsating stars or large amplitude eclipsing binaries. It can be run on ASCII and FITS light curve files, regardless of their origin.
Both the code and the processed light curves are publicly available, and we provide instructions for downloading and using them. The methodology used by \ksc will be applicable to future transit search missions such as TESS and PLATO. 
\end{abstract}


\section{Introduction}

NASA's \emph{Kepler} space mission brought about a revolution in high-precision, long baseline 
photometry, leading to major advances in exoplanet detection and characterization, 
asteroseismology, and the study of a wide range of stellar variability phenomena. After 
the failure of the second of its four reaction wheels in 2013, the original mission -- by 
then in its extended phase -- came to an end. The satellite was then re-purposed to survey 
four fields per year, located in or near the Ecliptic plane. This new mission, known as \Ktwo, 
delivers light curves of reduced photometric precision, owing to the reduced pointing 
accuracy, but offers unique opportunities to observe large numbers of bright ($V \le 12$) 
Sun-like stars, low-mass stars, and nearby open clusters, amongst other key targets 
\citep{how+14}.

\Ktwo data were initially released in the form of target pixel files (TPFs), which are 
time-series of individual postage stamps centered on each of the target stars. During the 
first year of operations, a number of teams developed pipelines to extract light curves 
from these TPFs \citep{vj14,van14,arm+14, arm+15,aig+15,hua+15,lib+15,buy+15}. Some of 
these are `all-purpose' in the sense that they can be applied to any TPFs, some are 
specifically designed for particular kinds of stars (e.g. asteroseismic targets) or for 
the `superstamps' made of many contiguous TPFs, which are used in \Ktwo to observe the 
crowded cores of dense open clusters. All of them include a light curve extraction step, 
which uses either aperture photometry (with circular or pixelated masks, whose positions 
can be static throughout an observation Campaign or adjusted to follow the star's position 
as the pointing varies) or point-spread function (PSF) fitting. All these pipelines also 
include a step to correct the systematic flux variations induced by the variations in the 
satellite's pointing (sometimes known as detrending). Almost without exception, the 
systematics correction in the aforementioned pipelines is done on an object-by-object 
basis, by modelling the dependence of the star's flux on its position. This represents a 
major departure from the approach most commonly used during the  \emph{Kepler} mission, as 
implemented in the \pdc\footnote{`PDC' stands for `pre-search data conditioning'.} and later 
\textsc{pdc-map}\footnote{`MAP' stands for `maximum a posteriori'.} components of the standard 
\emph{Kepler} pipeline, where `co-trending basis vectors' (CBVs) were constructed by 
combining many observed light curves, and each light curve was then decomposed into a 
linear combination of these CBVs plus a residual term including the star's intrinsic 
variations and random noise \textcolor{black}{\citep{smi+12,stu+12,stu+14,van+15}}.

From Campaign 3 onwards, the \Ktwo mission also started releasing light curve files (LCFs) 
extracted from the TPFs using a slightly modified version of the standard \emph{Kepler} Science Operations Center (KSOC)
pipeline. These LCFs include both the  simple aperture photometry (\sap) fluxes and the \pdc 
fluxes, but the latter typically contain significant residual systematics, indicating that 
the \pdc correction alone is not sufficient for \Ktwo. This motivated us to develop a somewhat 
improved and stand-alone version of the star-by-star systematics-correction step initially 
developed in \citet{aig+15} (hereafter Paper I), which can be applied to the \sap or \pdc light 
curves--but also, in principle, to the light curves produced by any of the 
aforementioned photometry pipelines. The present paper presents this systematics 
correction tool, which we call \ksc (\Ktwo Systematics Correction), including details of its 
implementation, photometric performance evaluation and signal injection and recovery 
tests, and links to the code (which is open source and freely available) as well as to all 
the \sap and \pdc data we have processed so far (all the individual long-cadence TPFs from 
Campaigns 3 to 5).

Our approach consists in modelling the observed flux as a Gaussian process (GP) with three 
additive components: the first depends on the star's position, and represents the 
pointing-induced systematics; the second depends on time, and represents the star's 
intrinsic variability as well as any other systematics not dependent on position; and the 
third represents white noise. The first component can then be subtracted to leave a light 
curve corrected for position-dependent systematics. The use of GP regression gives our 
model flexibility and robustness, and avoids having to specify an arbitrary functional 
form for the variations we seek to model. It is implemented in a  Bayesian  framework, 
which enables us to incorporate relevant physical information, for example in the form of priors on 
the length-scales of the various components, and to propagate the 
uncertainties associated with the systematics correction through to subsequent analysis. 
Our decision to model the star's intrinsic variations explicitly alongside the systematics 
was initially motivated by the desire to be able to correct systematics in variable stars 
without affecting the variability itself, but as we will show it also improves the 
systematics correction even in the case of relatively quiet stars. As a bonus, the 
time-component can also be subtracted to produce a light curve corrected for both 
systematics and stellar variability, which can be used, for example, to search for planetary 
transits.

Several improvements were made to our systematics correction method since Paper I. First, 
while we originally modelled the systematics as a function of the satellite roll-angle 
only, we now model them as a function of each star's two-dimensional position ($x$ and 
$y$). This gives better results for full \Ktwo Campaigns where the position variations 
include a significant drift as well as quasi-periodic roll-angle variations on $\sim 6$-hour timescales. Furthermore, the same roll-angle variations can correspond to very 
different actual changes in position at different locations in the field-of-view (FOV), so 
using the star's own position gives a better representation of the inter- and intra-pixel 
sensitivity variations which give rise to the systematics. We also introduced priors on 
the parameters of the GP model (length-scales and amplitudes for the various components) 
rather than performing a simple likelihood optimization within bounds. Finally, in 
addition to the standard (squared-exponential) kernel used to represent the stellar 
variability, which is suitable for smooth, aperiodic variations, we also implemented a 
quasi-periodic kernel which is better able to reproduce the behaviour of spotted stars and 
some classes of pulsators, along with a simple prescription for determining when this 
alternative kernel should be used.

Our method is described in detail in Section~\ref{sec:method}. We use two approaches to evaluate the performance of our
pipeline. The first is to measure the light curve scatter on transit
timescales, using a proxy measure of the Combined Differential Photometric
Precision (CDPP); these tests are reported in Section~\ref{sec:cdpp}. The second
is to inject simulated planetary transit signals into
the raw light curves and test our ability to recover them. Those tests are reported in Section~\ref{sec:injection_tests}. The \ksc package and data from Campaigns 3 to 5 are publicly available, and Section~\ref{sec:package} describes how to access and use them. We conclude and outline planes for future development in Section~\ref{sec:summary}.

\section{Method}
\label{sec:method}

\subsection{The basics}
\label{sec:basics}

We use GP regression to model the instrumental systematics, astrophysical variability and white noise in each light curve. A full description of GP regression is beyond the scope of this paper; we refer the interested reader to \citet{rw06} for a textbook introduction, and \citet{gib+12} and \citet{aig+12} for examples of GP regression applied to exoplanet datasets. 
Formally, a GP is a stochastic process, such that the joint probability distribution over any collection of observations of this process is a multi-variate Gaussian. The covariance matrix of this distribution is specified through a covariance function, which defines the covariance between pairs of observations as a function of a collection of input variables. This sets up a probability distribution over functions, which all share the desired covariance properties. Using well-known conditioning and marginalization identities for Gaussian distributions, it is then straight forward (and analytic) to condition this prior on the available data, and to evaluate the likelihood of the model, without having to specify a mathematical expression for the function directly. Standard optimization and model comparison methods can be used to fit for the parameters of the covariance function  and/or to compare different covariance functions. 
Importantly, if the covariance function contains several, additive terms, it is possible to separate the contribution of the different terms. We exploit this property here to separate the position-dependent systematics from the time-dependent astrophysical signal. 

\begin{figure*}
\centering
\includegraphics[width=\linewidth]{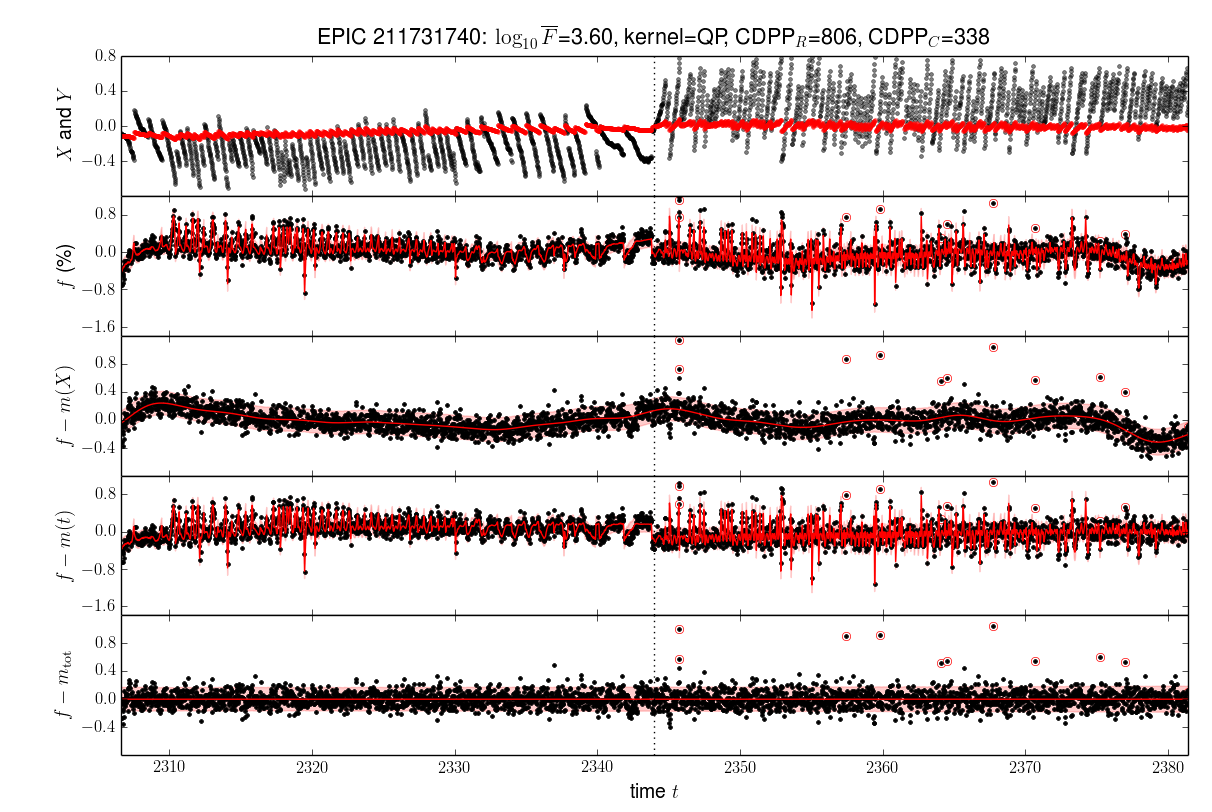}
\caption{Example Campaign 5 \Ktwo light curve before and after processing. From top to bottom: $x$ and $y$ position (black and red points, respectively), raw flux, flux corrected for systematics, flux corrected for intrinsic variations (showing systematics only), and residuals of the full model (after subtracting both the time and the position components). In the bottom four panels, black points show the data, the red curve shows the model, the pink shaded areas show the 95\% confidence interval about the model, and the points circled in red are identified as outliers and excluded from the model calculation. The vertical dotted line marks the point at which the direction of the roll-angle variations reverse, the position-dependent variations are treated separately either side of this line (see Section~\ref{sec:breakpoints}).}
\label{fig:example_time}
\end{figure*}

In the original version of our pipeline, described in Paper I, we used a single coordinate, representing the global roll-angle variations of the satellite, as the input driving the systematics component. Instead, the present version uses the 2-dimensional position of each star. When processing the KSOC light curves, our input variables representing position are the 
\verb\ POS_CORR1\ and \verb\POS_CORR2\ columns of the MAST light curves, which represent the deviation from the star's nominal position on the detector, 
as estimated by the KSOC pipeline from the satellite motion polynomials.

Each light curve is an array of flux measurements
$\flux = \{f_1,f_2,\ldots,f_n\}$. For simplicity, we will assume these have 
been normalized by dividing by the median and subtracting unity. The joint distribution over the flux measurement is assumed to be a multi-variate Gaussian with zero mean and covariance matrix \cmat:
\begin{equation}
\flux \sim \mathcal{N}(\mathbf{0},\cmat).
\end{equation}
The elements of the covariance matrix are given by:
\begin{align}
    \cmat_{ij} &= k(X_i,X_j) \nonumber \\
               &= k_{xy}(x_i,y_i,x_j,y_j) + k_t(t_i,t_j) + k_w(i,j)
\end{align}
where $X_i\equiv\{t_i,x_i,y_i\}$ and the three terms on the right-hand side represent, 
respectively, position-dependent systematics, the intrinsic variability of the star, and 
white noise. The position component uses a squared exponential covariance function 
with amplitude $A_{xy}$ and separate inverse length scales $\eta_x$ and $\eta_y$:
\[
k_{xy}(x_i,y_i,x_j,y_j) = A_{xy} \exp\left[ - \eta_x (x_i-x_j)^2 - \eta_y (y_i-y_j)^2 \right].
\]
This gives rise to smooth variations with characteristic amplitude
$A_{xy}$ and length-scales $1/\eta_x$ and $1/\eta_y$ in the $x$- and $y$-directions, respectively.
(We also tried using a single length-scale, i.e., a radial
covariance function, but obtained slightly better results with
separate length scales for $x$ and $y$.)
By default, the
time component also uses a squared exponential covariance function with
amplitude $A_t$ and inverse length scale $\eta_t$:
\[
k_{t}(t_i,t_j) = A_t \exp\left[ - \eta_t (t_i-t_j)^2 \right].
\]
This likewise gives smooth variations with characteristic amplitude
$A_t$ and time-scale $1/\eta_t$. This works well so long as
the star's intrinsic variability does not occur on time-scales similar
to the pointing variations. In Section~\ref{sec:variables} we
introduce a prescription for automatically switching to a more
adequate covariance function for periodic and quasi-periodic variables.
Finally, the white noise component is simply
\[
k_w(i,j) = \sigma^2 ~ \delta_{ij},
\]
where $\delta_{ij}$ is the Kronecker delta function.

The log-likelihood of the model is now
\[
\log p(\flux|\mathbf{\theta}) = -\frac{1}{2}  N\log(2\pi) -\frac{1}{2}\log(|\cmat|) - 
\flux^T \cmat^{-1} \flux,
\]
where $\mathbf{\theta} \equiv {A_{xy}, \eta_x, \eta_y, A_t, \eta_t, \sigma}$ are the 
parameters of the covariance function, $|K|$ is the determinant of the covariance matrix, 
and $\flux^T$ is the transpose of $\flux$. The GP kernels and log-likelihood are 
implemented using the {\sc george} package\footnote{See {\tt http://dan.iel.fm/george}.} \citep{amb+14}.

\begin{figure}
\centering
\includegraphics[width=\linewidth]{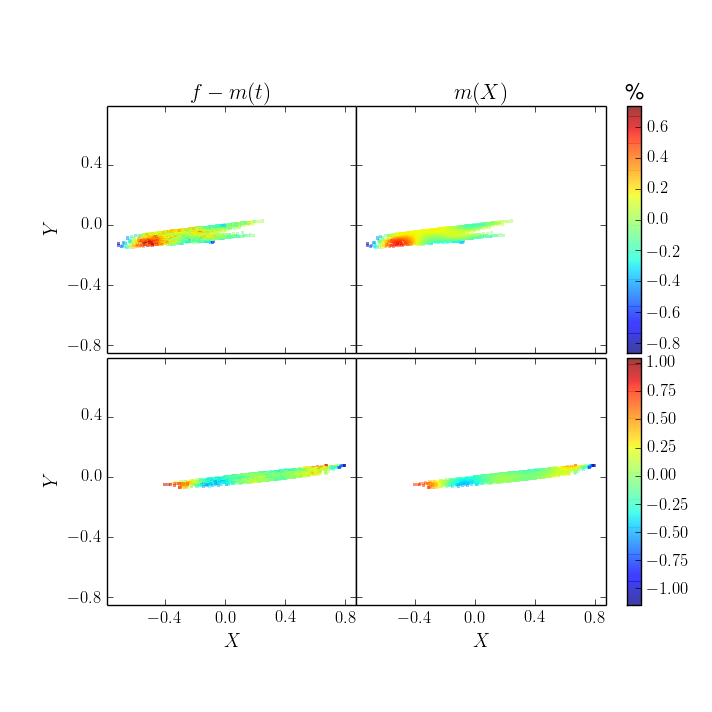}
\caption{Position-dependence of the flux after removing the intrinsic variations (left) and the position-dependent component of the model (right), for the light curve shown in Figure~\ref{fig:example_time}. The top and bottom panels show data taken before and after the break-point (dotted vertical line on Figure~\ref{fig:example_time}. The data and model agree closely, and the differences between the behaviour before and after the break-point are noticeable. Note also that the variations are clearly two-dimensional, so a 1-D model for the position dependence would not work as well. Finally, in the bottom row we see a hint of periodicity in the $x$-axis, with a period of 1 pixel, which is what one would expect if the main source of systematics is to intra-pixel sensitivity variations.}
\label{fig:example_pos}
\end{figure}

We define priors over each of the parameters of the covariance function:
\begin{itemize}
\item $A_{xy}$ and $A_t$: uniform prior between -7 and 1 in normalized $\log_{10}$ flux 
      (that is, flux divided by its median);
\item $\sigma$: uniform prior between -6 and 0 in normalized $\log_{10}$ flux;
\item $\eta_x$ and $\eta_y$: truncated normal prior with mean 17 pixel$^{-1}$, standard 
      deviation 8~pixel$^{-1}$, minimum 0 pixel$^{-1}$, and maximum 70 pixel$^{-1}$; 
\item $\eta_t$: truncated log-normal prior with mean 0.25 day$^{-1}$, standard 
      deviation 1.25 dex, and upper boundary at 2 day$^{-1}$ to 
      prevent over-fitting the noise.
\end{itemize}
We learn the GP hyperparameters from the data by finding the posterior density maximum 
using a global optimization step followed by local optimization. First, we use a 
differential evolution (DE) global optimizer \citep{sp97} adapted from \textsc{PyDE}\footnote{\url{https://github.com/hpparvi/PyDE},  DOI:10.5281/zenodo.45602} \citep{Parviainen2016DE} to explore the full range of parameter 
space and generate a population of parameter vectors clumped close to the global posterior 
maximum. The DE starting population is initialized to uniformly sample a subvolume of the 
parameter space allowed by the priors (this is because the uniform prior high boundaries 
are much higher than what would be expected for any realistic case), but the population is 
allowed to expand to explore outside the initial boundaries. After the DE step, the 
parameter vector with the highest posterior value is chosen as a starting point for local 
optimization using the Powell's local optimization method \citep[][as 
implemented in SciPy]{powell1964}.

For a given set of covariance parameters, the predictive
distribution for the model conditioned on the data, evaluated at any
set of inputs $X_*=\{t_*, x_*, y_*\}$, is a Gaussian with mean
\[
\overline{f}_* = \cmat_* \cmat^{-1} \flux
\]
and variance
\[
{\rm var}(f_*) = \cmat_{**} - \cmat_* \cmat^{-1} \cmat_*^T
\]
where $\cmat_* \equiv \{ k(X_*,X_1), k(X_*,X_2), \ldots, k(X_*,X_n)\}$ and
$\cmat_{**} \equiv k(X_*,X_*)$. Importantly, we can evaluate the
predictive mean for the different components of the model separately
\[
\overline{f}_{*,t} = \cmat_{*,t} \cmat^{-1} \flux
\]
where $\cmat_{*,t} \equiv \{ k_t(t_*,t_1), k_t(t_*,t_2), \ldots,
k_t(t_*,t_n)\}$, and similarly
\[
\overline{f}_{*,xy} = \cmat_{*,xy} \cmat^{-1} \flux
\]
where $\cmat_{*,xy} \equiv \{ k_{xy}(x_*,y_*,x_1,y_1), k_{xy}(x_*,y_*,x_2,y_2), \ldots $, 
$ k_{xy}(x_*,y_*,x_n,y_n) \}$. 

We can thus remove the
position-dependent systematics from the raw light curve, while
preserving intrinsic variability, by evaluating $\overline{f}_{*,xy}$
at each cadence, and subtracting it from the original. Furthermore, it
is also straightforward to evaluate the variance of the predictive
mean for the position component alone, and hence to obtain a robust
uncertainty estimate for each point in the corrected light curve. 

We also compute and store the full model prediction ($\overline{f}_*$
evaluated at each cadence), as this can be subtracted from the raw
data to obtain a version of the light curve where both systematics and
the smooth component of any intrinsic variability have been removed:
this can in principle be useful in searching for short-duration events
such as eclipses, planetary transits and flares.

Figures~\ref{fig:example_time} and \ref{fig:example_pos} show examples of the light curve before and after processing, and the time- and position-dependence of the model, for a typical star observed in \Ktwo Campaign~5. In this case, the `raw' light curve was the \sap flux. When working with LCFs produced by the standard \emph{Kepler} pipeline, we use the position of the star predicted from the motion polynomials of the satellite (stored in the \verb\POS_CORR1\ and \verb\POS_CORR2\ columns), which should be more precise and less sensitive to contamination effects than the centroid-based estimates (stored in the \verb\MOM_CENTR1\ and \verb\MOM_CENTR2\ columns).

\subsection{Outlier detection}
\label{sec:outlier}

If left untreated, outliers can be a significant problem: the model seeks to explain them by adopting an unphysically-short length-scale for the time component. To address this, we use an initial outlier detection step. We carry out an initial detrending using a default set of hyperparameters based on the complete hyperparameter population of Campaign~4 \textcolor{black}{(see Section~\ref{sec:hp} for a discussion of the hyper-parameter distributions)}, and flag any points deviating more than 5$\sigma$ above or below of the detrended light curve as outliers. The kernel hyperparameter optimization is then carried out using a subset of the 
remaining points, and the predictive distribution is computed again using the best-fit parameters. Points located more 
than 5$\sigma$ above or below this prediction are again flagged, and a new, final prediction is generated. This final 
prediction uses only points, which were not flagged as outliers, but it is computed for all data points, including 
outliers. For example, deep transits and stellar eclipses are often flagged as outliers, but it is still desirable to 
compute a corrected light curve including these events.

\begin{figure*}
\centering
\includegraphics[width=\linewidth]{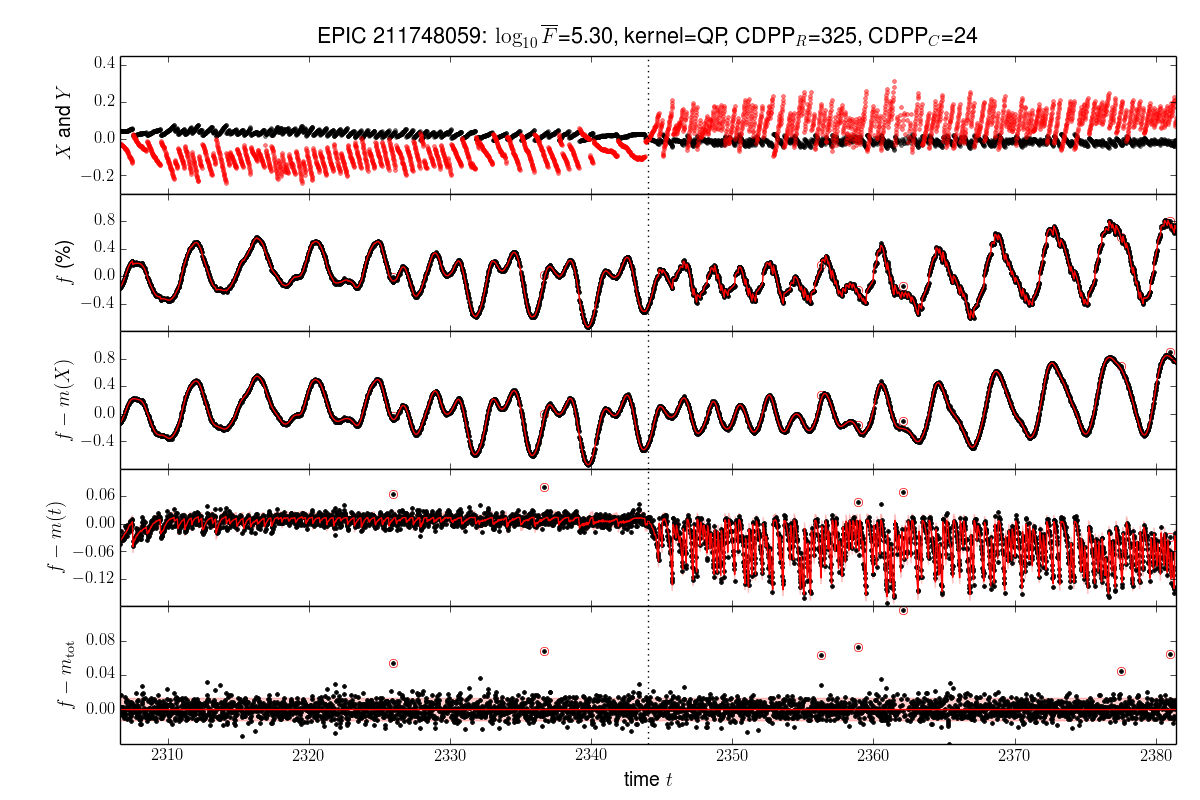}
\caption{Same as Figure~\ref{fig:example_time}, but for a light curve displaying quasi-periodic variations.}
\label{fig:example_qp}
\end{figure*}

\begin{figure}
\centering
\includegraphics[width=\linewidth]{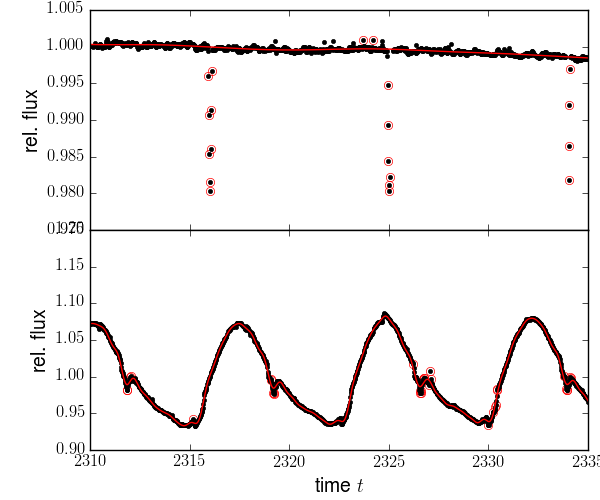}
\caption{\textcolor{black}{Segments of the systematics-corrected PDC light curves for two eclipsing binaries from \Ktwo Campaign 5, illustrating the successful (top) and unsuccessful (bottom) identification of in-eclipse points as outliers. The symbols and lines are the same as for Fig.~\ref{fig:example_time}.}}
\label{fig:outl_examples}
\end{figure}

\textcolor{black}{Despite the relatively sophisticated outlier treatment described above, eclipses are not always flagged as outliers. Specifically, if there is significant out-of-eclipse variability on timescales similar to the eclipse duration, the length scale of the time-component converges to a value that is short compared to the eclipse duration. Consequently, the time-component of the model attempts to fit the eclipses as well as the out-of-eclipse variability, but generally struggles to reproduce the sharp ingress and egress of the eclipses. To illustrate this, Figure~\ref{fig:outl_examples} shows light curve segments for two eclipsing binaries from \Ktwo Campaign 5. The first has short-duration eclipses and little out-of-eclipse variability, and the eclipses are duly flagged as outliers. The second has significant variability and longer eclipses, and the in-eclipse points are flagged as outliers. In such a case, not only would removing the time-component of the model remove most of the eclipse signal, but the systematics correction is also less satisfactory during the eclipses. To address this issue, we included an option in \ksc to manually force in-eclipse points to be treated as outliers for indvidual objects, by specifying the ephemeris and eclipse duration.}

\subsection{Break-points}
\label{sec:breakpoints}

Visual examination of \Ktwo light curves before and after modelling them using the method described above soon revealed 
that the behaviour of the systematics changes qualitatively at one or two points during each campaign. This is clearly visible in the light curve shown in Figure~\ref{fig:example_qp}, for example (which also illustrates how we deal with periodic variable stars, as discussed in Section~\ref{sec:variables}).
These points correspond to reversals of the direction of the roll-angle variations: the net torque due to solar radiation pressure pushes the satellite first one way, then the other as the campaign progresses and the orientation of \emph{Kepler} changes relative to the Sun. It is not entirely clear why this leads to different behaviours in the systematics, it may be due to small changes in the star's position during each exposure (30 minutes for long-cadence data). We modified the position-dependent term in the covariance function to include a break-point each time the roll-angle variations change direction. This is implemented by multiplying the original covariance matrix with a binary mask which takes the value of unity if both points belong to the same segment (between two reversals of the roll-angle drift) and zero otherwise. The break-points between segments are the same for all objects in a given \Ktwo campaign, and are determined by visual inspection of the position variations for a few dozen light curves spanning the FOV. We note that changing the precise timing of the break-points (up to a day or two) does not affect the results significantly.

\subsection{Handling variable stars}
\label{sec:variables}

The model described in Section~\ref{sec:basics} performs well for stars with moderate intrinsic variability occurring on timescales 
considerably longer than the characteristic timescale of the \Ktwo roll angle variations (approx.\ 6 hours). In those 
cases, the two components of the GP adequately separate the intrinsic variability from the position-dependent 
systematics. However, the correction is less successful for strong variables such as classical pulsators or active, rapidly rotating 
stars. In those cases, as already noted in Paper I, $A_t$ and $A_{xy}$ tend to shrink to zero, and both systematics and 
intrinsic variability are typically absorbed into the white noise, which becomes much larger than normal (for a given 
star brightness).  The correction is thus `conservative', meaning that it does not remove true variability, but it also 
leaves much of the systematics unaffected.

In Paper I, we noted that this problem could be overcome on a case-by-case basis by altering the initial guess for 
$\eta_t$, or by implementing a more explicit model of the variability. This would be suitable for relatively rare kinds 
of variables, such as pulsating stars. However, one of the strengths of \Ktwo is its ability to observe young open 
clusters, whose members are typically active and rapidly rotating. These are too numerous to be treated manually. We 
have therefore implemented an automated procedure for identifying and handling variable stars that display a clear 
periodicity. First, we compute the Lomb-Scargle periodogram of the raw light curve \citep{Lomb1976,Press} in the 
period range 0.05--20 days. If the false alarm probability of the periodigram maximum is lower than a given threshold value (by default $10^{-50}$), we 
replace the time-component of the GP model with the following, quasi-periodic covariance function:
\[
k_{\rm rot} (t_i,t_j) = A_t \exp\left[ -
  \Gamma \sin^2\left(\frac{\pi |t_i-t_j|}{P}\right) - \frac{(t_i-t_j)^2}{L^2_e}  \right],
\]
where $P$ is the period, $\Gamma$ the inverse length scale of the periodic component of the variations, and $L_e$ the 
evolutionary time-scale of the variations. $P$ is initially set to the period of the periodogram peak, and the 
evolutionary time-scale to 10 times that value. The covariance parameters are then refined using the same procedure as 
for the non-periodic case. This gives significantly improved results for spotted stars and pulsating stars with 
pulsation periods of a day or more. As a by-product, this procedure yields estimates of the dominant period of the 
stellar variability and of its characteristic evolutionary timescale, which are stored in the headers of the corrected 
light curve files. Figure~\ref{fig:example_qp} shows an example light curve for a spotted star before and after 
correction.

\subsection{Hyper-parameter distributions}
\label{sec:hp}

\begin{figure*}
\centering
\includegraphics[width=0.48\linewidth]{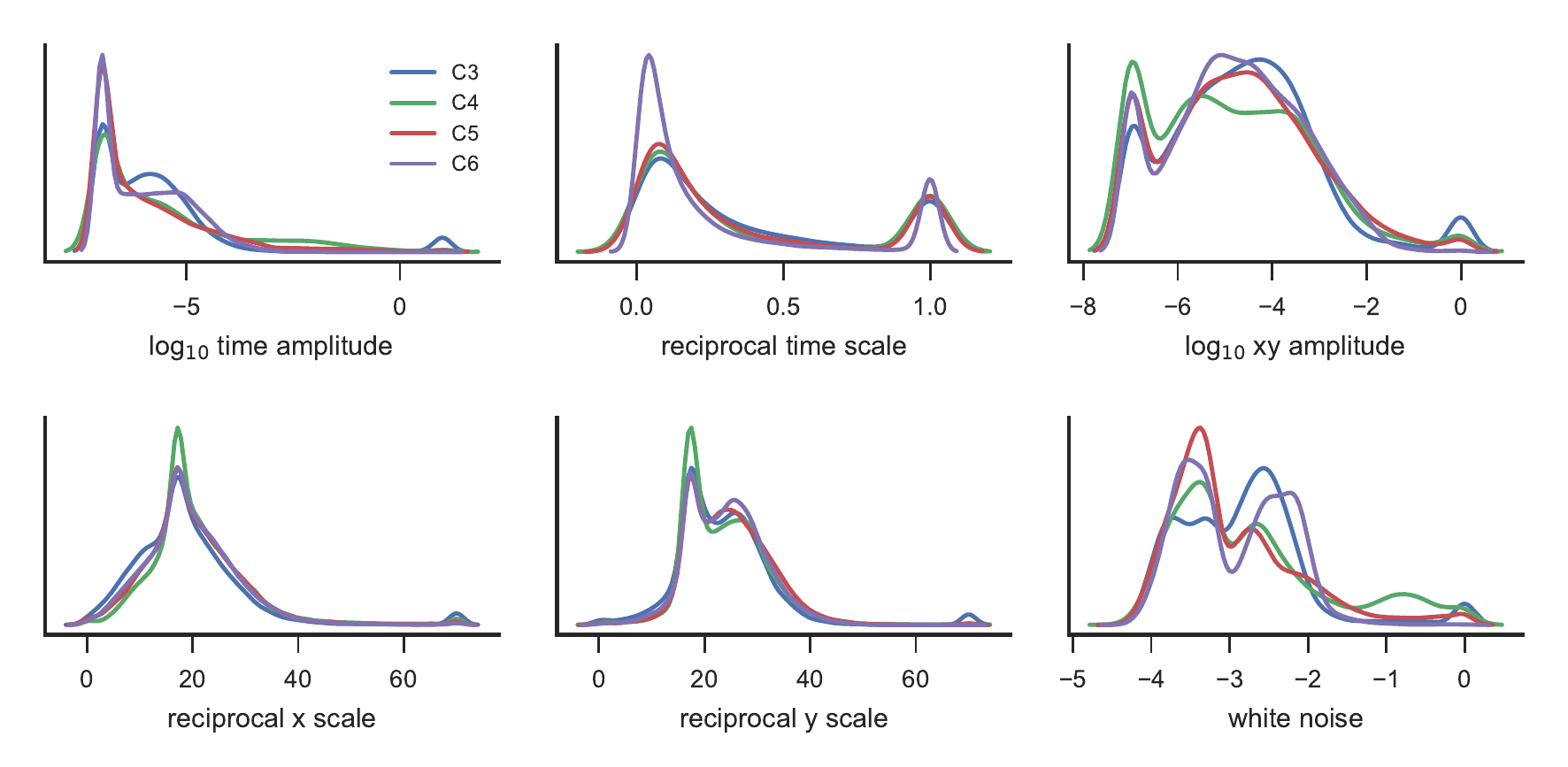}
\hfill
\includegraphics[width=0.48\linewidth]{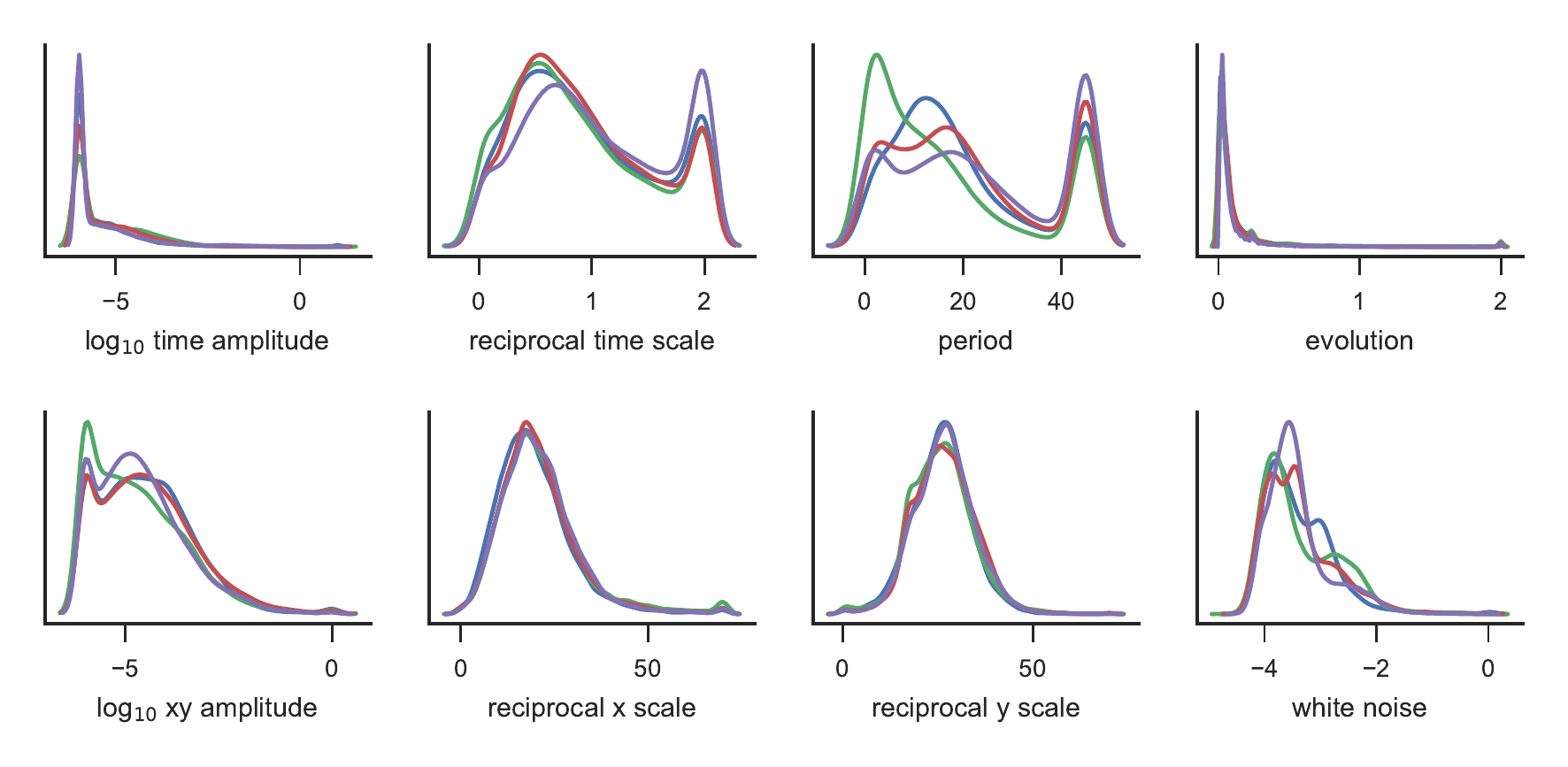}
\caption{\textcolor{black}{Final hyper-parameter distributions for Campaigns 3 to 6, for the non-periodic cases (left) and the quasi-periodic cases (right).}}
\label{fig:hp}
\end{figure*}

\textcolor{black}{Figure~\ref{fig:hp} shows the distributions of the final (best-fit) hyper-parameters for Campaigns 3 to 6, for the non-periodic and quasi-periodic cases, respectively. The distributions are broadly consistent between the different campaigns, which implies that the noise, systematics and variability properties of the light curves do not change significantly from one campaign to the next. This provides an a-posteriori justification of our choice to use the medians of the distributions from Campaign 4 (which was the first campaign we processed in full) as the default values for the initial detrending performed prior to identifying outliers (see Section~\ref{sec:outlier}). }

\textcolor{black}{The distributions for the white noise term show more variation between campaigns, but this is due to the different magnitude distributions of the targets. In the quasi-periodic cases, the final periods do not necessarily match the initial guess taken from the Lomb-Seeliger results. The secondary peak in the period distribution at around 30 days, corresponds to cases where the period and evolutionary timescales are similar. In such cases, the model reverts to a random, rather than clearly periodic behaviour. Focusing on periods below 30 days, we also note that the distribution for Campaign 4 peaks at shorter periods, which may be a result of the larger fraction of young (Pleiades and Hyades) stars observed in this campaign.}

\textcolor{black}{We also examined the hyper-parameter distributions as a function of location in the FOV, and noted that the hyper-parameters associated with the systematics component of the model depend somewhat on distance from the satellite boresight. Incorporating this information in the initial guesses may improve the detrending, and will be considered in the future.}

\section{Photometric performance for \Ktwo data}
\label{sec:cdpp}

\begin{figure}
\centering
\includegraphics[width=\linewidth]{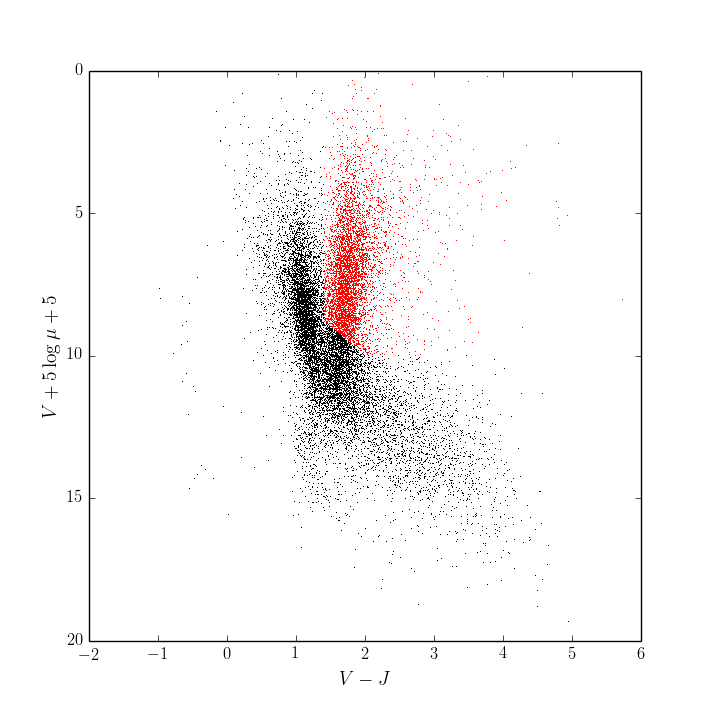}
\caption{Reduced proper motion diagram for \Ktwo Campaign~5 targets. The black points follow the locus of Galactic disk main sequence stars. The red points, corresponding to stars with red colours and low proper motion for their magnitude, are likely red giants.}
\label{fig:reduced_pm}
\end{figure}

\begin{figure*}
\centering
\includegraphics[width=\linewidth]{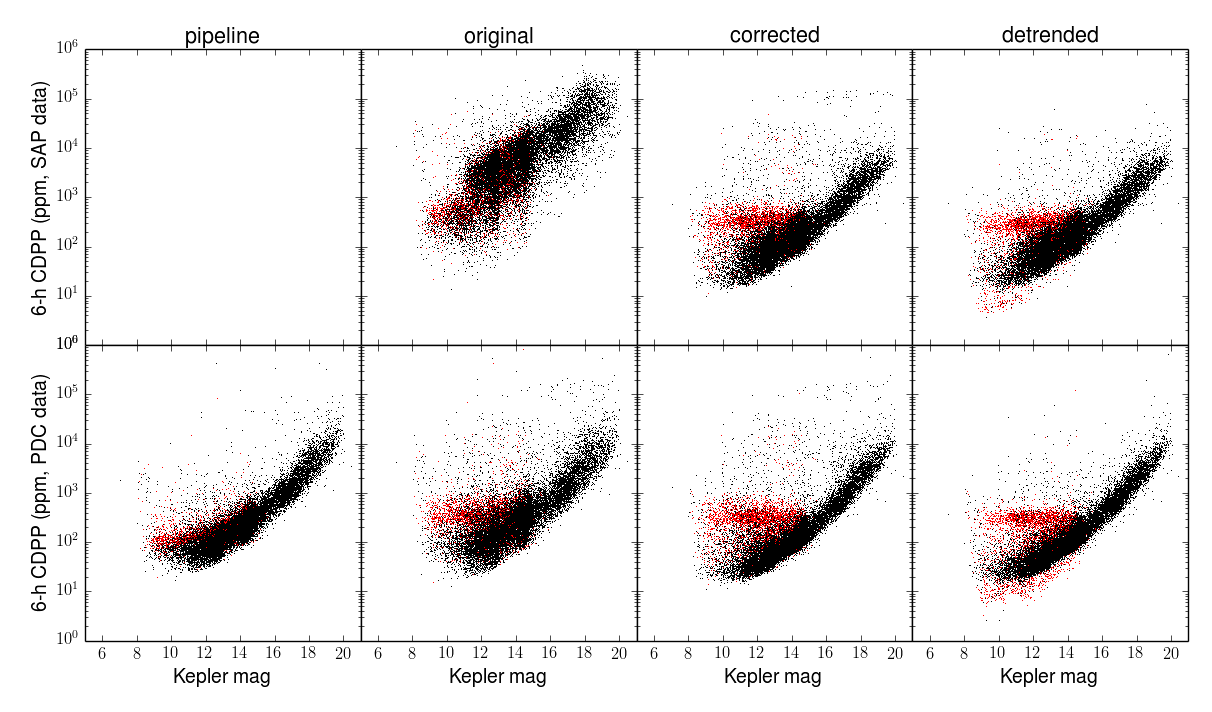}
\caption{6.5-hour CDPP versus \emph{Kepler} magnitude for \Ktwo Campaign 5, starting from the \sap or the \pdc light curves (top and bottom, respectively). From left to right: \emph{Kepler} pipeline CDPP estimates (these are available for \pdc data only), and our CDPP estimates for the original data, the data corrected for systematics, and the data corrected for systematics and stellar variability. Objects identified in the reduced proper motion diagram (Figure~\ref{fig:reduced_pm}) as likely giants are shown in red.}
\label{fig:cdpp_all}
\end{figure*}

The CDPP has become a de-facto standard for evaluating the photometric precision of \emph{Kepler} and \Ktwo light curves. It is formally defined  as the inverse signal-to-noise ratio (SNR) of a reference transit signal of the corresponding duration, in parts per million (ppm), and can be interpreted as the depth of a transit of the given duration to exhibit an SNR of 1 \citep{chr+12}. The KSOC transit search pipeline  systematically evaluates the CDPP on
3, 6.5 and 12-hour timescales. Since we do not have access to this pipeline, we can
only evaluate a proxy measurement of the CDPP. To do this, we follow
the same procedure as \citet{gil+11}: trends on timescales longer than 2 days are removed
using a Savitzky-Golay filter, then 5-$\sigma$ clipping is used to
remove large outliers. We then compute the average flux in consecutive
13-point bins, keeping only the bins with $>7$ valid data points. The
6-hour CDPP proxy is then estimated as the standard deviation of the
bin-averaged fluxes. For each light curve processed with \ksc, this evaluation is performed on the input light curve, the light curve after removing systematics (i.e. subtracting the position-dependent component of the GP model), and the light curve after removing both systematics and stellar variability (i.e. subtracting also the time-dependent component of the GP model).

\begin{figure}
\centering
\includegraphics[width=\linewidth]{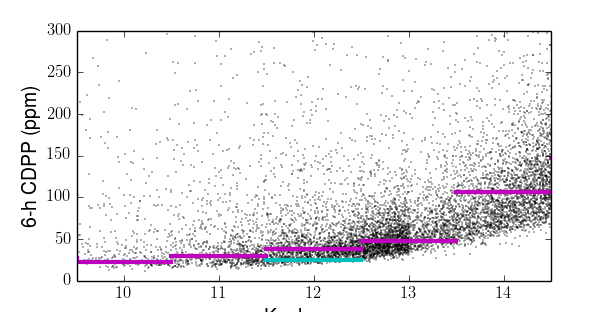}
\caption{6.5-hour CDPP versus \emph{Kepler} magnitude for \Ktwo Campaign 5, excluding the giants and focussing on the brighter part of the magnitude range. The horizonal magenta lines show the median CDPP in 1-magnitude bins. For comparison, the cyan line shows the median CDPP obtained for stars with $11.5 \le Kp \le 12.5$ observed during the original \emph{Kepler} mission \protect\citep{gil+11}.}
\label{fig:cdpp_bright}
\end{figure}

Figure~\ref{fig:cdpp_all} shows the 6-hour CDPP proxy for all light curves available from MAST for \Ktwo Campaign 5 (data release 7, does not include superstamps or short-cadence targets). The data are shown as a function of \emph{Kepler} magnitude ($Kp$), as provided in the EPIC catalog \citep{hub+16_epic}. Only objects for which the $Kp$ value included in the EPIC catalog was derived from the Sloan $g$, $r$ and $i$ magnitudes are shown (as opposed to values derived from other magnitude systems such as $B$ and $V$ or 2MASS $JHK$). In the first column, we show for reference the 6-hour CDPP values estimated by the KSOC pipeline (these are available for the \pdc data only). The remaining three column show our own CDPP proxy estimates, for the `raw', systematics-corrected and systematics- and variability-corrected data ($2^{\rm nd}$, $3^{\rm rd}$ and $4^{\rm th}$ columns, respectively). Note that we have also processed the data for Campaigns 3 and 4, and the results are similar.

The first and second panels in the bottom rows should look similar if our CDPP proxy is accurate; it does for the majority of the points, with the notable exception of those shown in red. These are objects identified as likely giants in a reduced proper motion diagram (see Figure~\ref{fig:reduced_pm}), and are discussed further in Section~\ref{sec:giants}. Our CDPP proxy estimates for the \sap data are considerably higher and more scattered, as expected since the \pdc step partially removes both the \Ktwo systematics and stellar variability on timescales of days to weeks \citep{smi+12,stu+12,stu+14}. 

Giants aside, the CDPP proxy values for systematics-corrected data ($3^{\rm rd}$ column) are considerably lower on average, and follow a tighter relation with $Kp$ magnitude, than those for the raw light curves. This shows that our \ksc pipeline yields a significant improvement over the basic data products, and holds whether the \sap or the \pdc data are used as the starting point, though the \pdc light curves yield slightly lower CDPP values. Removing the time component of our model (final column) reduces the number of outliers lying above the main relation (variable stars) but otherwise does not change the diagram significantly.

In Figure~\ref{fig:cdpp_bright}, we take a closer look at the photometric performance for dwarf stars with $Kp \le 14.5$. The lower enveloppe of the distribution ranges from $<20$\,ppm (for $Kp \sim 10$) to about 70\,ppm (for $Kp \sim 14$, which is comparable with the performance recorded during the original \emph{Kepler} mission. This implies that, for a significant fraction of the stars, we are able to remove the additional systematics caused by the roll-angle variations of the satellite to a level where they are below other sources of noise. We have also computed the median CDPP values in 1-magnitude bins (magenta lines). For the bin $11.5 \le Kp \le 12.5$, the median value is 38\,ppm. This is only 50\% higher than the value of 25\,ppm median CDPP derived in exactly the same manner, for the same magnitude range, for light curves from the original \emph{Kepler} mission \citep{gil+11}.

\begin{figure*}
\centering
\includegraphics[width=\textwidth]{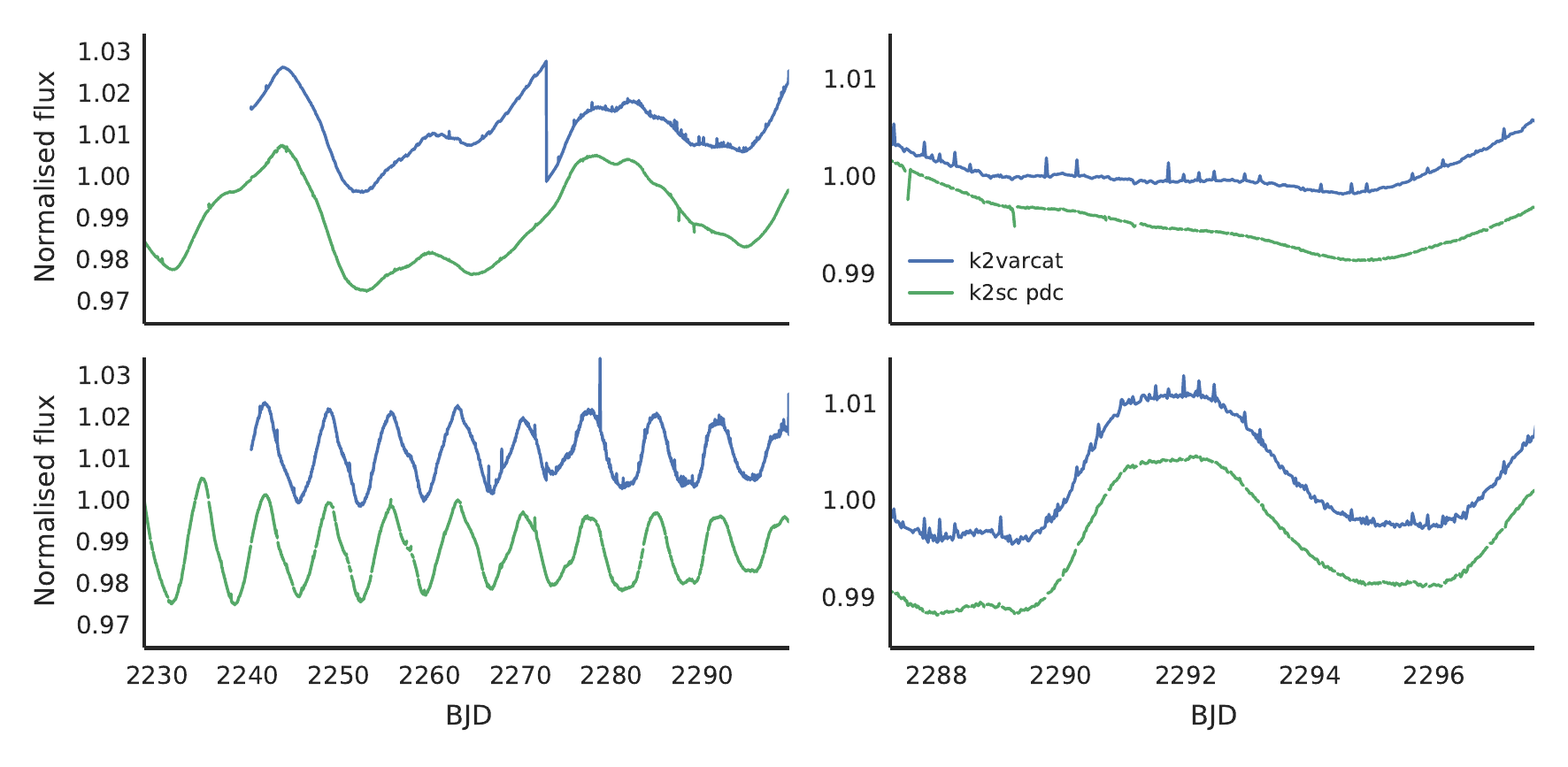}
\caption{Comparison between the \kvc light curves (blue) and the \ksc-detrended \sap light curves (green, vertically offset for clarity) for four Campaign 4 objects. The full Campaign 4 duration is shown on the left, and a zoomed subset on the right.}
\label{fig:comp_warwick}
\end{figure*}

\subsection{Giant stars}
\label{sec:giants}

Red giants display excess variability on few hours timescales due to stochastically excited pulsations. The KSOC transit-search pipeline includes a harmonic-filtering step which is designed to filter out these oscillations \citep{jen+10,ten+12}, and hence yields lower CDPP estimates for giants than our proxy values, which are based on the \pdc data without any filtering. The same applies to the results after removing the systematics, but once the time component is also subtracted, a small fraction of the red points move below the main envelope of the relation. This can be understood as follows. The time component of the model for giants tends to use a quasi-periodic kernel with a short period, which fits the oscillations, but if this period is too close to the Nyquist frequency of the data the time-component also explains some of the white (random) noise. On the other hand, in some cases the Lomb-Scargle periodogram criterion for using a quasi-periodic GP is not met, and the variability of the giants is not modelled at all, so there is still a cluster of red points above the main envelope in the final column of Figure~\ref{fig:cdpp_all}. Better performance for pulsating stars could be obtained by modelling each oscillation frequency using  e.g.\ a cosine kernel GP, but this would require an object-by-object treatment, which is beyond the scope of the present paper.

\subsection{Comparison to other pipelines}
\label{sec:comp}

\begin{figure}
\centering
\includegraphics[width=\columnwidth]{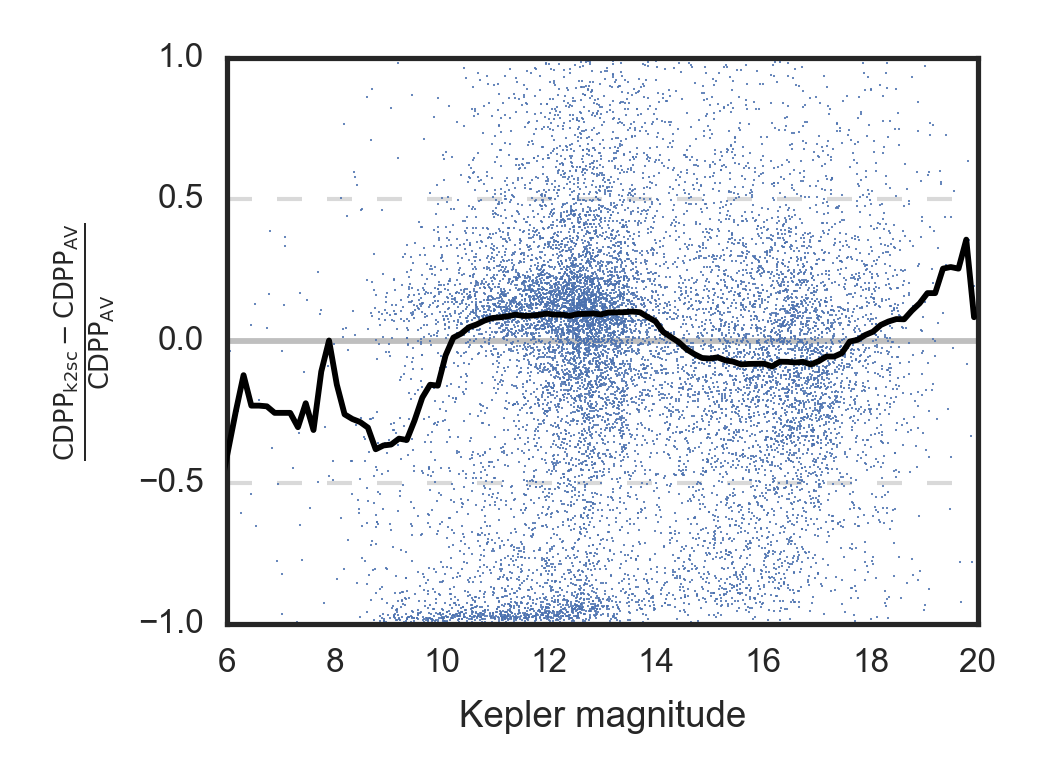}
\caption{The relative CDPP difference between the \ksf  and \ksc-corrected \pdc light curves for Campaign 4. The blue dots show the individual differences, and the black line the running median with a width of 0.5 mag. The clump of points with relative CDPP difference close to 01 are red giants (see discussion in Section~\ref{sec:giants}).}
\label{fig:comp_relative}
\end{figure}

A number of other teams have developped their own pipelines to extract and detrend \Ktwo light curves \citep[see e.g.]{vj14,van14,hua+15,k2p2,arm+15,arm+16}. It would not be practical to include here a detailed comparison to all of these -- added to which, not all of the resulting light curves are public, or available for the campaigns which we have processed with \ksc. We therefore focus our comparison on the \ksf \citep{vj14,van14} and \kvc \citep{arm+15,arm+16} pipelines, as the light curves produced by these pipelines are available at the time of writing, as `high level science products' (HLSPs) on the MAST \Ktwo archive, for campaigns 1 to 4.

Applying our \ksc pipeline to the \sap or \pdc light curves systematically gives better results than the \kvc pipeline. This is illustrated for two example cases from campaign 4, on Figure~\ref{fig:comp_warwick}. The examples include both variable and quiet stars; in either case both pipelies give broadly similar results, but the \ksc light curves are considerably less noisy. In the first example, the \kvc also contains a large discontinuity, which is not present in the \ksc version. We note that, in some cases, the raw \kvc light curves are less noisy, and show less obvious systematics, than the \sap or \pdc light curves, perhaps because the apertures used are more optimal. In those cases, applying \ksc to the raw \kvc might give even better results, but this is outside the scope of the present paper. 

We also compare the \ksc-detrended \pdc light curves from campaign 4 to the output of the widely used \ksf pipeline. Figure~\ref{fig:comp_relative} shows the relative difference in CDPP between the two sets of light curves. In both cases, we computed the proxy CDPP estimates ourselves using the procedure described above, as the procedure described in \cite{vj14} to estimate CDPP differs slightly from ours (specifically, it tends to give slightly lower values if any correlated noise on transit timescales remains). Except at the end of the magnitude range, where there are too few objects to make a meaningful comparison, the relative difference in CDPP is below 15\% on average. The \ksf light curves tend to have slightly lower CDPP for bright stars ($Kp \le 14$) and slightly higher CDPP for fainter stars. (The same pattern was already seen in Paper I using data from the \Ktwo engineering test). However, across the entire magnitude range, there are cases where the CDPP values are very different in either direction. This implies that it may be useful to compare the \ksf and \ksc-corrected light curves for any individual object, before making a decision on which to use. Once again, even better results may be obtained by applying \ksc to the raw \ksf light curves, but this is not possible at present, as the publicly available \ksf light curve files do not include 2-D position information.

\section{Transit injection tests}
\label{sec:injection_tests}

We tested how the \ksc detrending affects transit searches by carrying out injection tests using 7000 randomly selected Campaign 5 stars. In each case, we simulated a transit signal using PyTransit\footnote{Available at \url{https://github.com/hpparvi/PyTransit}.} \citep{Parviainen2015pt}, and injected it into the \pdc light curve. We then detrended the light curve using \ksc, and ran a basic transit search using our own implementation\footnote{Available at \url{https://github.com/hpparvi/PyBLS}.} of the box least-squares (BLS) algorithm \citep{Parviainen2016BLS} of \citet{kov+02}. We also ran the BLS search for the data with only partial detrending, where we removed the \ksc time component (variability) but did not remove the position component (systematics). This partially detrended dataset mimics a normal transit search case with basic time-based detrending, and works as a baseline against which we test the effects of the full detrending. Finally, we carried out the detrending and  BLS search for the same dataset, but without transit signal injection to identify any existing signals present in the original light curves. This led us to remove $\sim$50 stars from the analysis because they had a strong BLS detection statistic (signal detection efficiency ${\rm SDE}>10$) without an injected signal.

\textcolor{black}{As the PDC-MAP pipeline is not publicly available, we could not inject the simulated transit signals into the light curves before the PDC-MAP step. Therefore, the tests described in this section do not account for the effects of the PDC-MAP pipeline on transit signals, but only for the effect of our \ksc pipeline on such signals. However, we note the PDC-MAP pipeline is designed to preserve transit signals, if necessary at the expense of other astrophysical signals. For a detailed investigation of the effect of the PDC-MAP pipeline on simulated transit signals, see  \citet{2012PASP..124.1279C}.}

The transit signals were simulated with a sampling of 3\,min, before integrating them to the \Ktwo cadence of 30\,min. The parameters of the simulated transits were drawn from the following distributions:
\begin{itemize}
\item stellar density $\rho_\star$: Gaussian with mean $1.7$ and standard deviation $\sigma=0.1$\,g\,cm$^{-3}$;
\item orbital period $P$: uniform from 0.75 to 40\,days;
\item planet-to-star radius ratio $k$: half-Gaussian with mean 0 and s$\sigma=0.75$ (positive half only).
\end{itemize}
The impact parameter and orbital eccentricity were set to zero for simplicity, and we fixed the quadratic limb darkening law coefficients to 0.4 and 0.1.

The BLS search was carried out over a uniform grid of 5000 frequencies from 0.025 to 1.33\,day$^{-1}$ (0.75 to 40 days) with \textcolor{black}{$q_{\rm min}$ of 0.001, $q_{\rm max}$ of 0.2 (where $q_{\rm min}$ and $q_{\rm max}$ are the minimum and maximum transit durations in units of the orbital period, respectively)} and per-frequency binning of 500, and the signal giving the highest SDE was recorded as the detected signal. 
The injected signal was considered as correctly identified if the recovered period was within 2\% of the injected period. The BLS algorithm computes a periodogram of signal-to-noise ratios (SNRs) of the best fit transit at each trial period. The detection statistic used is then the signal detection efficiency (SDE), which is the ratio of the highest peak in the SNR periodogram to its standard deviation. 

\begin{figure}
\centering
\includegraphics[width=\columnwidth]{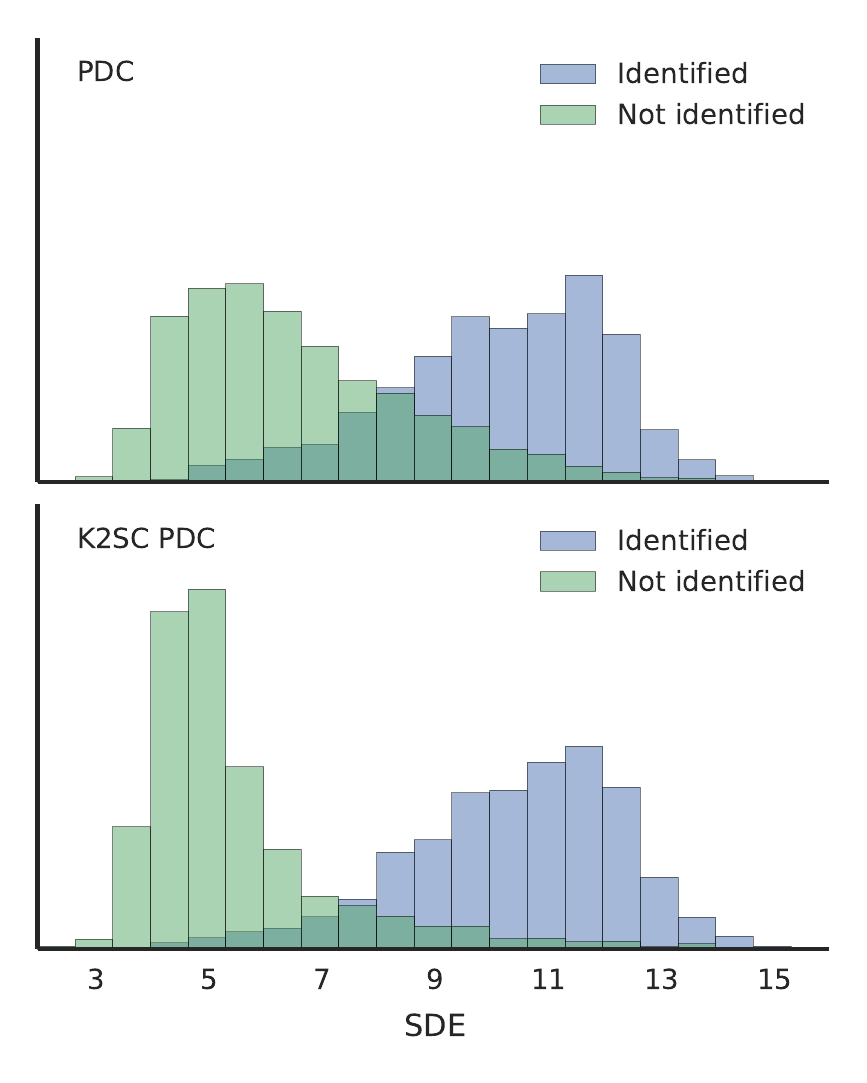}
\caption{BLS signal detection efficiency (SDE) for the identified and unidentified injected signals for original \pdc data (upper panel) and the \ksc-detrended \pdc data (bottom panel).}
\label{fig:inj_sde}
\end{figure}

\begin{figure}
\centering
\includegraphics[width=\columnwidth]{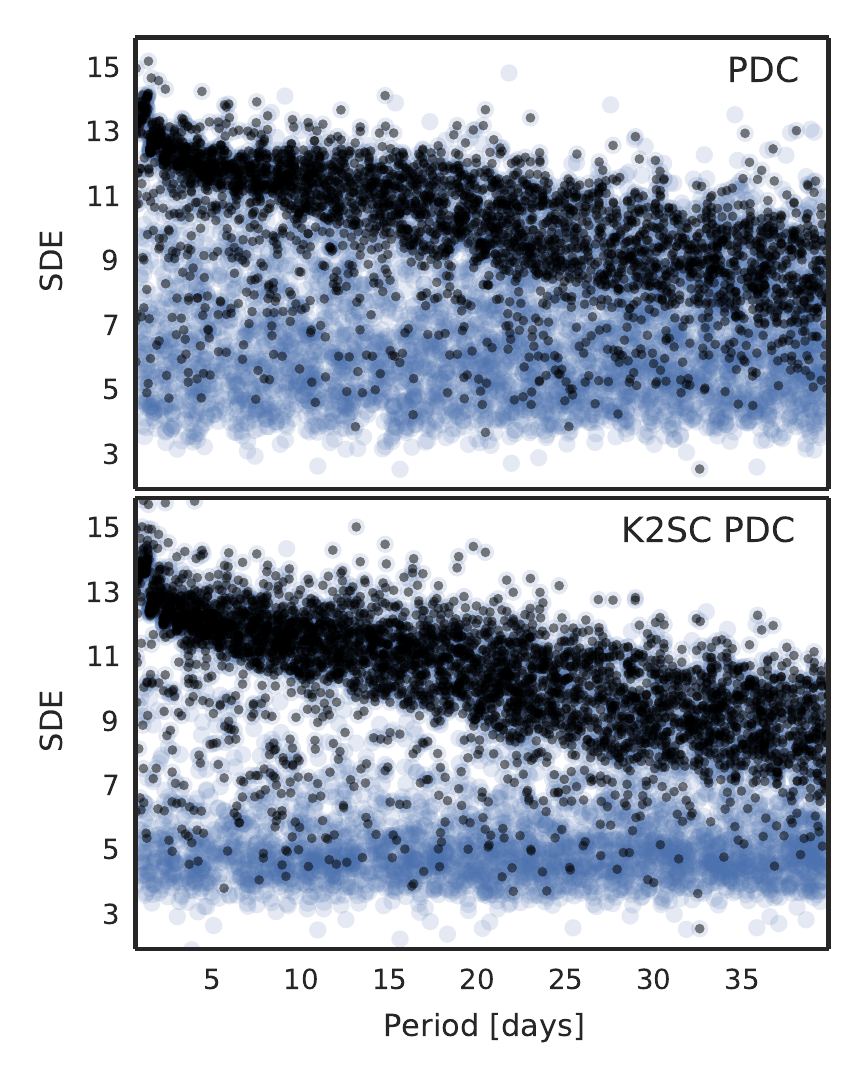}
\caption{BLS signal detection efficiency (SDE) for the identified and unidentified signals as a function of period of the injected signal for the original (upper panel) and \ksc-detrended (lower panel) \pdc data. Blue dots mark all the injected signals, and black dots the correctly identified signals.}
\label{fig:inj_sde_vs_p}
\end{figure}

\begin{figure*}
\centering
\includegraphics[width=\textwidth]{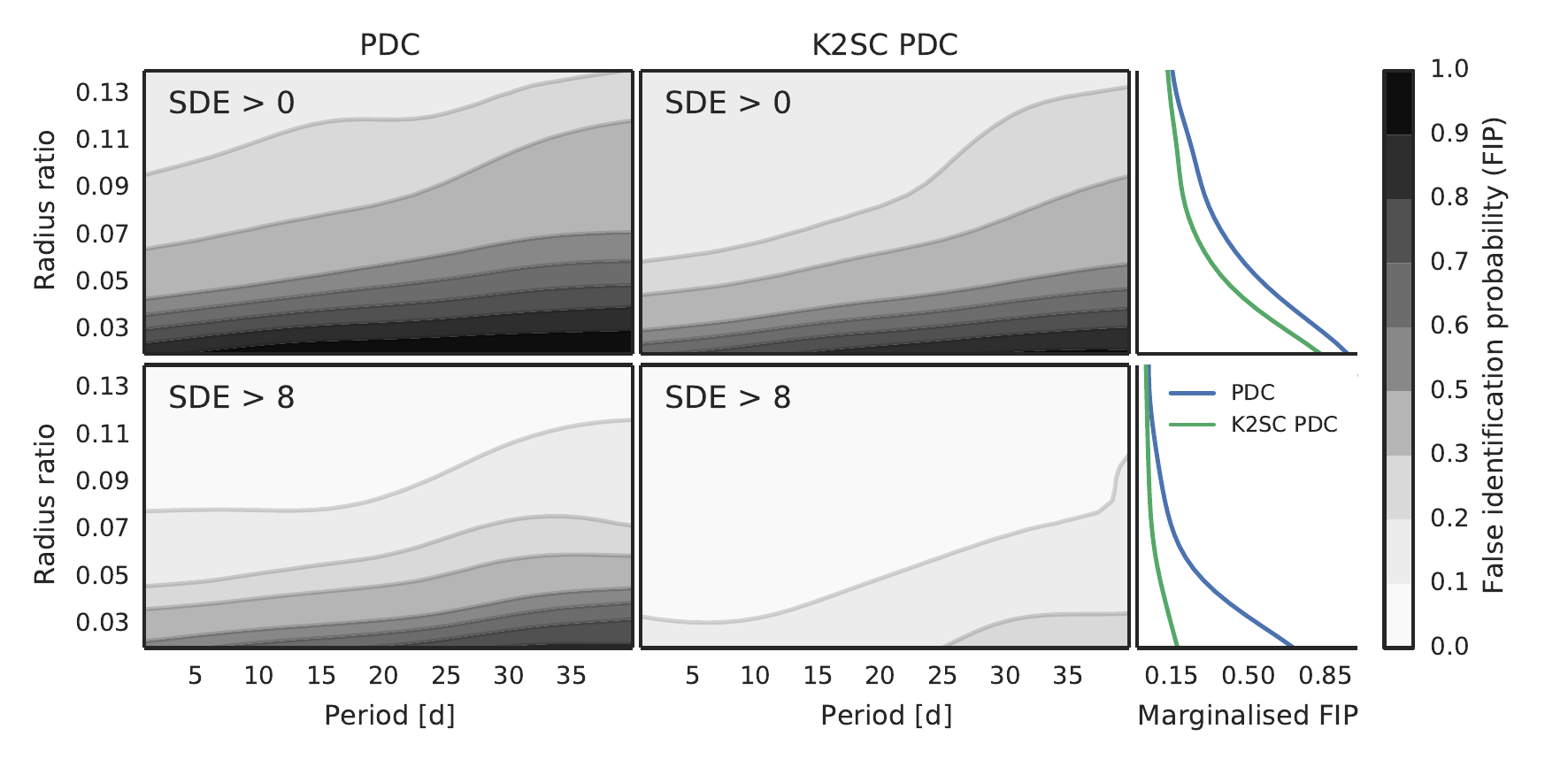}
\caption{The false identification probability (FIP) as a function of radius ratio and orbital period for the whole dataset (upper row), and for signals with BLS SDE$>$8 (lower row). The first column shows the FIP for the original \pdc data, the second column shows the FIP for the \ksc-detrended \pdc data, and the third column shows the FIP marginalised over the orbital period. The probability densities are calculated using Gaussian kernel density estimation. The completeness of the search can be calculated as 1-FIP for SDE$>$0.}
\label{fig:fap_comparison}
\end{figure*}

Since all the light curves contained an injected signal, but none of them had strong BLS signal before injection, the cases where the recovered signal was not the injected one (which we refer to as `unidentified signals') are cases where another signal was detected instead. i.e.\ false alarms. Figure \ref{fig:inj_sde} shows the BLS SDE distributions for the correctly identified and unidentified signals for the original and \ksc-detrended \pdc data, and Fig.~\ref{fig:inj_sde_vs_p} shows the same as a function of the injected orbital period.

Figure \ref{fig:fap_comparison} shows the false identification probability (FIP) as a function of radius ratio and orbital period for original and detrended \pdc data, as well as the FIP marginalized over the injected period. The FIP is defined as the fraction of the detected signals which are not correctly identified. 
The upper row shows the FPI for the whole dataset, without imposing any detection threshold. The lower row shows a more realistic transit search case where we used a detection threshold of ${\rm SDE} \ge 8$. (In an even more more realistic case, one might use a variable SDE cut as a function of identified radius ratio and period.)
The \ksc detrending improves our capability to correctly identify the injected transit signals over the whole $(p,k)$-space independent of SDE value, as shown from the first row. More importantly, when applying an SDE cut, we see a significant difference in our ability to recover injected signals, particularly those corresponding to small planets (radius ratio smaller than 0.05).

\section{Publicly available code and data}
\label{sec:public}

\subsection{The \protect\ksc package}
\label{sec:package}

\ksc is distributed as a Python package containing the main \ksc executable and a set of Python modules. The executable can be used to detrend \Ktwo light curves from several sources, and offers an automatic MPI parallelization for detrending large datasets using computing clusters. The \ksc Python modules offer a lower-level access to the code, which is useful when, for example, using a custom \Ktwo photometry pipeline and the light curve format is not supported by the executable, or when the user wants to experiment with new GP kernels. 

The package is open source with a GPLv3 license, and is available from GitHub
\begin{quote}
  \url{htpps://github.com/OxES/k2sc}
\end{quote}
and from the Python Package Index (PyPi). The installation follows normal Python package installation steps, and we have aimed to keep the external dependencies in minimum (SciPy, NumPy, George, and PyFITS\footnote{PyFITS is a product of the Space Telescope Science Institute, which is operated by AURA for NASA.}). The code has been written keeping an eye on extendability, and adding new GP kernels and importers for data from custom photometry pipelines is easy.

The detrending time for a single light curve on a modern computer is dominated by the given number of differential evolution iterations to carry out before the local hyperparameter optimization. Generally, the results do not improve significantly after 1-2 minutes of hyperparameter optimization, which leads to a total detrending time of $\sim$2 minutes per light curve.  

\subsection{Processed light curves from \Ktwo campaigns 3 to 5}
\label{sec:data}

We have processed all the light curves generated using the KSOC pipeline and available at the Mikulski Archive for Space Telescopes (MAST). At the time of writing, this includes the long-cadence light curves from Campaigns 3 to 5. 
 These \ksc-processed LCs are available from MAST as a \Ktwo High-Level Science Product (HLSP). For more information on how to search for, download and use the \ksc HLSPs, see \textcolor{black}{{\tt archive.stsci.edu/prepds/k2sc/.}}
 
\section{Conclusions and future work}
\label{sec:summary}

We have presented a new pipeline, \ksc, to model and, if desired, remove instrumental systematics and astrophysical variability in light curves from space-based transit surveys. The development of such a pipeline was motivated by the strong roll-induced systematics present in \Ktwo data, but the same method could in principle be applied many other datasets. The \ksc pipeline models the systematics as a GP depending on the star's position on the detector, and the astrophysical variability as a separate, additive GP depending on time only. It automatically checks for periodic behaviour in the light curves and, of this is found, adapts the time component accordingly. We have applied this \ksc to the publicly available long-cadence \Ktwo \sap and \pdc light curves, and evaluated the photometric precision of the resulting detrended light curves by computing a proxy estimate of their CDPP on 6-hour timescales. The CDPP for bright stars ($V\sim12$) is within a factor 1.5 of the original \emph{Kepler} mission. 

We compared our results to publicly available light curves produced by other pipelines, specifically \kvc \citep{arm+15,arm+16} and \ksf \citep{vj14,van14}. Our light curves are consistently less noisy and contain fewer `glitches' and discontinuities than the \kvc ones. The precision we achieve (measured in terms of 6-hour CDPP) is broadly similar to that of the \ksf pipeline, but the latter gives slightly better results for bright stars ($V \le 14$) while \ksc gives slightly better results for fainter stars. We speculate that this may be due to differences between the apertures used by the KSOC pipeline (which produces the \sap and \pdc light curves, to which we applied \ksc) and the \ksf pipeline. An important advantage of \ksc over \ksf is that it is more robust to astrophysical variability. Specifically, because the systematics and variability are modelled simultaneously, \ksc is better able to distinguish between the two, even when the variability is strong. Furthermore, our GP-based approach to modelling variability, including where appropriate a  kernel that reproduces the quasi-periodic behaviour of rotating active stars, combined with a careful treatment of outliers, enables the variability to be subtracted to reveal even low-amplitude transits and other short-lived events.
While \ksc is designed to be general-purpose, we therefore recommend it particularly to anyone who is interested in studying variable stars with \Ktwo, or in searching for transits or flares in the light curves of variable stars.

We also used signal injection and recovery tests to explore the completeness of planetary transit searches in \ksc-detrended \pdc light curves, for planets with radius ratio $k$ down to 0.01 and period $P$ up to 40\,d in some cases. We demonstrated better than 50\% sensitivity down to $k=0.03$ at $P=5$\,d, rising to 0.05 at $P=40$\,d (see Figure~\ref{fig:fap_comparison}, upper middle panel). This implies that \Ktwo should be able to detect warm Neptunes and hot super-Earths around Sun-like stars, and some Earth-sized planets in the habitable zones of M dwarfs.

While the results to date are very promising, the photometric performance might be improved further, for example by using a different set of light curves than the \sap or \pdc as input. The \ksc pipeline can in principle by applied to any light curve so long as it contains time, flux and 2-D position information. For example, light curves extracted with more optimal apertures might yield better end results. We note that users wishing to read a new light curve format with \ksc can readily do so by writing a small wrapper function, or are welcome to contact us for help if required. On a separate note, the \ksc-processed \Ktwo light curves we have examined, as well as light curves produced by other pipelines, often display some long-term trends, which are common to many stars. Such trends were also common in the original \emph{Kepler} data, they can be caused by a range of factors from aperture losses associated with the long-term drift of the telescope pointing to temperature and focus changes. The \pdc pipeline was designed to remove such trends from \sap light curves, but it may function less effectively in the presence of the large amplitude systematics associated with the \Ktwo roll-angle variations. It might be more effective to model the latter first using \ksc, and then attempt to remove the common-mode trends using a principal component analysis (PCA)-like method, such as e.g.\ SysRem, \citep{tam+05} or the slightly more sophisticated ARC method we developed for the original \emph{Kepler} mission \citep{rob+13}.

\textcolor{black}{Another possible avenue for further improvement might be improving the initial guesses for the GP hyper-parameters, for example incorporating information about the dependence of the position-dependent systematics component of our model on an object's location in the \emph{Kepler} FOV.}

For users interested in a particular kind of variable stars, better results might be achieved on an object-by-object basis by using a specialized GP covariance function to model the time-component in the \ksc framework. For example, for pulsating stars, one might use a sum of decaying cosine kernels (as employed by \citealt{bs09}), with priors on the pulsating frequencies derived from a preliminary analysis of the light curve power spectrum. Once again, adding custom-built GP kernels to \ksc should be relatively easy, and prospective users are encouraged to try it themselves and/or contact us for help. 

Finally, we conclude by noting that the methodology we have developed here is readily applicable to other datasets, potentially including ground-based transit surveys such as the Next Generation Transit Survey (NGTS), as well as the upcoming Transiting Exoplanet Survey Satellite (TESS) and the future PLATO mission. Since these instruments may suffer from different kinds of systematics, it might be necessary to modify the systematics component of the model, potentially including other input variables (e.g.\ detector temperature, seeing, airmass, and so on\ldots) but the overall framework should still apply. 

\section*{Acknowledgments}

We wish to thank Andrew Vanderburg, Daniel Foreman-Mackey, Ann Marie Cody and Tom Barclay for useful discussions and feedback.
This work was made possible by financial support from the Leverhulme Trust (RPG-2012-661), the UK Science and Technology Facilities Council (ST/K00106X/1) and the Clarendon Trust. The data presented in this paper were obtained from the Mikulski Archive for Space Telescopes (MAST). STScI is operated by the Association of Universities for Research in Astronomy, Inc., under NASA contract NAS5-26555. Support for MAST for non-HST data is provided by the NASA Office of Space Science via grant NNX09AF08G and by other grants and contracts.

\bibliography{k2sc}

\begin{thebibliography}{}

\bibitem[\protect\citeauthoryear{{Aigrain}, {Hodgkin}, {Irwin}, {Lewis} \&
  {Roberts}}{{Aigrain} et~al.}{2015}]{aig+15}
{Aigrain} S.,  {Hodgkin} S.~T.,  {Irwin} M.~J.,  {Lewis} J.~R.,    {Roberts}
  S.~J.,  2015, MNRAS, 447, 2880

\bibitem[\protect\citeauthoryear{{Aigrain}, {Pont} \& {Zucker}}{{Aigrain}
  et~al.}{2012}]{aig+12}
{Aigrain} S.,  {Pont} F.,    {Zucker} S.,  2012, MNRAS, 419, 3147

\bibitem[\protect\citeauthoryear{{Ambikasaran}, {Foreman-Mackey}, {Greengard},
  {Hogg} \& {O'Neil}}{{Ambikasaran} et~al.}{2014}]{amb+14}
{Ambikasaran} S.,  {Foreman-Mackey} D.,  {Greengard} L.,  {Hogg} D.~W.,
  {O'Neil} M.,  2014, ArXiv e-prints

\bibitem[\protect\citeauthoryear{{Armstrong}, {Kirk}, {Lam}, {McCormac},
  {Osborn}, {Spake}, {Walker}, {Brown}, {Kristiansen}, {Pollacco}, {West} \&
  {Wheatley}}{{Armstrong} et~al.}{2016}]{arm+16}
{Armstrong} D.~J.,  {Kirk} J.,  {Lam} K.~W.~F.,  {McCormac} J.,  {Osborn}
  H.~P.,  {Spake} J.,  {Walker} S.,  {Brown} D.~J.~A.,  {Kristiansen} M.~H.,
  {Pollacco} D.,  {West} R.,    {Wheatley} P.~J.,  2016, MNRAS, 456, 2260

\bibitem[\protect\citeauthoryear{{Armstrong}, {Kirk}, {Lam}, {McCormac},
  {Walker}, {Brown}, {Osborn}, {Pollacco} \& {Spake}}{{Armstrong}
  et~al.}{2015}]{arm+15}
{Armstrong} D.~J.,  {Kirk} J.,  {Lam} K.~W.~F.,  {McCormac} J.,  {Walker}
  S.~R.,  {Brown} D.~J.~A.,  {Osborn} H.~P.,  {Pollacco} D.~L.,    {Spake} J.,
  2015, A\&A, 579, A19

\bibitem[\protect\citeauthoryear{{Armstrong}, {Osborn}, {Brown}, {Kirk}, {Lam},
  {Pollacco}, {Spake} \& {Walker}}{{Armstrong} et~al.}{2014}]{arm+14}
{Armstrong} D.~J.,  {Osborn} H.~P.,  {Brown} D.~J.~A.,  {Kirk} J.,  {Lam}
  K.~W.~F.,  {Pollacco} D.~L.,  {Spake} J.,    {Walker} S.~R.,  2014, ArXiv
  e-prints

\bibitem[\protect\citeauthoryear{{Brewer} \& {Stello}}{{Brewer} \&
  {Stello}}{2009}]{bs09}
{Brewer} B.~J.,  {Stello} D.,  2009, MNRAS, 395, 2226

\bibitem[\protect\citeauthoryear{{Buysschaert}, {Aerts}, {Bloemen},
  {Debosscher}, {Neiner}, {Briquet}, {Vos}, {P{\'a}pics}, {Manick}, {Schmid},
  {Van Winckel} \& {Tkachenko}}{{Buysschaert} et~al.}{2015}]{buy+15}
{Buysschaert} B.,  {Aerts} C.,  {Bloemen} S.,  {Debosscher} J.,  {Neiner} C.,
  {Briquet} M.,  {Vos} J.,  {P{\'a}pics} P.~I.,  {Manick} R.,  {Schmid} V.~S.,
  {Van Winckel} H.,    {Tkachenko} A.,  2015, MNRAS, 453, 89

\bibitem[\protect\citeauthoryear{{Christiansen}, {Jenkins}, {Caldwell},
  {Burke}, {Tenenbaum}, {Seader}, {Thompson}, {Barclay}, {Clarke}, {Li},
  {Smith}, {Stumpe}, {Twicken} \& {Van Cleve}}{{Christiansen}
  et~al.}{2012a}]{chr+12}
{Christiansen} J.~L.,  {Jenkins} J.~M.,  {Caldwell} D.~A.,  {Burke} C.~J.,
  {Tenenbaum} P.,  {Seader} S.,  {Thompson} S.~E.,  {Barclay} T.~S.,  {Clarke}
  B.~D.,  {Li} J.,  {Smith} J.~C.,  {Stumpe} M.~C.,  {Twicken} J.~D.,    {Van
  Cleve} J.,  2012a, PASP, 124, 1279

\bibitem[\protect\citeauthoryear{{Christiansen}, {Jenkins}, {Caldwell},
  {Burke}, {Tenenbaum}, {Seader}, {Thompson}, {Barclay}, {Clarke}, {Li},
  {Smith}, {Stumpe}, {Twicken} \& {Van Cleve}}{{Christiansen}
  et~al.}{2012b}]{2012PASP..124.1279C}
{Christiansen} J.~L.,  {Jenkins} J.~M.,  {Caldwell} D.~A.,  {Burke} C.~J.,
  {Tenenbaum} P.,  {Seader} S.,  {Thompson} S.~E.,  {Barclay} T.~S.,  {Clarke}
  B.~D.,  {Li} J.,  {Smith} J.~C.,  {Stumpe} M.~C.,  {Twicken} J.~D.,    {Van
  Cleve} J.,  2012b, PASP, 124, 1279

\bibitem[\protect\citeauthoryear{{Gibson}, {Aigrain}, {Roberts}, {Evans},
  {Osborne} \& {Pont}}{{Gibson} et~al.}{2012}]{gib+12}
{Gibson} N.~P.,  {Aigrain} S.,  {Roberts} S.,  {Evans} T.~M.,  {Osborne} M.,
  {Pont} F.,  2012, MNRAS, 419, 2683

\bibitem[\protect\citeauthoryear{{Gilliland}, {Chaplin}, {Dunham},
  {Argabright}, {Borucki}, {Basri}, {Bryson}, {Buzasi}, {Caldwell}, {Elsworth},
  {Jenkins}, {Koch}, {Kolodziejczak}, {Miglio}, {van Cleve}, {Walkowicz} \&
  {Welsh}}{{Gilliland} et~al.}{2011}]{gil+11}
{Gilliland} R.~L.,  {Chaplin} W.~J.,  {Dunham} E.~W.,  {Argabright} V.~S.,
  {Borucki} W.~J.,  {Basri} G.,  {Bryson} S.~T.,  {Buzasi} D.~L.,  {Caldwell}
  D.~A.,  {Elsworth} Y.~P.,  {Jenkins} J.~M.,  {Koch} D.~G.,  {Kolodziejczak}
  J.,  {Miglio} A.,  {van Cleve} J.,  {Walkowicz} L.~M.,    {Welsh} W.~F.,
  2011, APjS, 197, 6

\bibitem[\protect\citeauthoryear{{Howell}, {Sobeck}, {Haas}, {Still},
  {Barclay}, {Mullally}, {Troeltzsch}, {Aigrain}, {Bryson}, {Caldwell},
  {Chaplin}, {Cochran}, {Huber}, {Marcy}, {Miglio}, {Najita}, {Smith},
  {Twicken} \& {Fortney}}{{Howell} et~al.}{2014}]{how+14}
{Howell} S.~B.,  {Sobeck} C.,  {Haas} M.,  {Still} M.,  {Barclay} T.,
  {Mullally} F.,  {Troeltzsch} J.,  {Aigrain} S.,  {Bryson} S.~T.,  {Caldwell}
  D.,  {Chaplin} W.~J.,  {Cochran} W.~D.,  {Huber} D.,  {Marcy} G.~W.,
  {Miglio} A.,  {Najita} J.~R.,  {Smith} M.,  {Twicken} J.~D.,    {Fortney}
  J.~J.,  2014, PASP, 126, 398

\bibitem[\protect\citeauthoryear{{Huang}, {Penev}, {Hartman}, {Bakos},
  {Bhatti}, {Domsa} \& {de Val-Borro}}{{Huang} et~al.}{2015}]{hua+15}
{Huang} C.~X.,  {Penev} K.,  {Hartman} J.~D.,  {Bakos} G.~{\'A}.,  {Bhatti} W.,
   {Domsa} I.,    {de Val-Borro} M.,  2015, MNRAS, 454, 4159

\bibitem[\protect\citeauthoryear{{Huber}, {Bryson}, {Haas}, {Barclay},
  {Howell}, {Sharma}, {Stello} \& {Thompson}}{{Huber}
  et~al.}{2016}]{hub+16_epic}
{Huber} D.,  {Bryson} S.~T.,  {Haas} M.~R.,  {Barclay} T.,  {Howell} S.~B.,
  {Sharma} S.,  {Stello} D.,    {Thompson} S.~E.,  2016, ApJS, submitted,
  arXiv:1512.02643

\bibitem[\protect\citeauthoryear{{Jenkins}, {Caldwell}, {Chandrasekaran},
  {Twicken}, {Bryson}, {Quintana}, {Clarke}, {Li}, {Allen}, {Tenenbaum}, {Wu},
  {Klaus}, {Van Cleve}, {Dotson}, {Haas}, {Gilliland}, {Koch} \&
  {Borucki}}{{Jenkins} et~al.}{2010}]{jen+10}
{Jenkins} J.~M.,  {Caldwell} D.~A.,  {Chandrasekaran} H.,  {Twicken} J.~D.,
  {Bryson} S.~T.,  {Quintana} E.~V.,  {Clarke} B.~D.,  {Li} J.,  {Allen} C.,
  {Tenenbaum} P.,  {Wu} H.,  {Klaus} T.~C.,  {Van Cleve} J.,  {Dotson} J.~A.,
  {Haas} M.~R.,  {Gilliland} R.~L.,  {Koch} D.~G.,    {Borucki} W.~J.,  2010,
  ApJL, 713, L120

\bibitem[\protect\citeauthoryear{{Kov{\'a}cs}, {Zucker} \&
  {Mazeh}}{{Kov{\'a}cs} et~al.}{2002}]{kov+02}
{Kov{\'a}cs} G.,  {Zucker} S.,    {Mazeh} T.,  2002, A\&A, 391, 369

\bibitem[\protect\citeauthoryear{{Libralato}, {Bedin}, {Nardiello} \&
  {Piotto}}{{Libralato} et~al.}{2015}]{lib+15}
{Libralato} M.,  {Bedin} L.~R.,  {Nardiello} D.,    {Piotto} G.,  2015, MNRAS,
  in press, arXiv:1510.09180

\bibitem[\protect\citeauthoryear{Lomb}{Lomb}{1976}]{Lomb1976}
Lomb N.~R.,  1976, Astrophysics and Space Science, 39, 447

\bibitem[\protect\citeauthoryear{{Lund}, {Handberg}, {Davies}, {Chaplin} \&
  {Jones}}{{Lund} et~al.}{2015}]{k2p2}
{Lund} M.~N.,  {Handberg} R.,  {Davies} G.~R.,  {Chaplin} W.~J.,    {Jones}
  C.~D.,  2015, ApJ, 806, 30

\bibitem[\protect\citeauthoryear{Parviainen}{Parviainen}{2015}]{Parviainen2015pt}
Parviainen H.,  2015, MNRAS, 450, 3233

\bibitem[\protect\citeauthoryear{Parviainen}{Parviainen}{2016a}]{Parviainen2016BLS}
Parviainen H., , 2016a, PyBLS: v0.9

\bibitem[\protect\citeauthoryear{Parviainen}{Parviainen}{2016b}]{Parviainen2016DE}
Parviainen H., , 2016b, PyDE: v1.5

\bibitem[\protect\citeauthoryear{{Powell}}{{Powell}}{1964}]{powell1964}
{Powell} M.~J.~D.,  1964, The Computer Journal, 7, 155

\bibitem[\protect\citeauthoryear{Press, Teukolsky, Vetterling \&
  Flannery}{Press et~al.}{2007}]{Press}
Press W.,  Teukolsky S.,  Vetterling W.,    Flannery B.,  2007, {Numerical
  Recipes: The Art of Scientific Computing}, 3 edn.
Cambridge University Press, Cambridge

\bibitem[\protect\citeauthoryear{{Rassmussen} \& {Williams}}{{Rassmussen} \&
  {Williams}}{2006}]{rw06}
{Rassmussen} C.~E.,  {Williams} C. K.~I.,  2006, Gaussian Processes for Machine
  Learning.
The MIT Press

\bibitem[\protect\citeauthoryear{{Roberts}, {McQuillan}, {Reece} \&
  {Aigrain}}{{Roberts} et~al.}{2013}]{rob+13}
{Roberts} S.,  {McQuillan} A.,  {Reece} S.,    {Aigrain} S.,  2013, MNRAS, 435,
  3639

\bibitem[\protect\citeauthoryear{{Smith}, {Stumpe}, {Van Cleve}, {Jenkins},
  {Barclay}, {Fanelli}, {Girouard}, {Kolodziejczak}, {McCauliff}, {Morris} \&
  {Twicken}}{{Smith} et~al.}{2012}]{smi+12}
{Smith} J.~C.,  {Stumpe} M.~C.,  {Van Cleve} J.~E.,  {Jenkins} J.~M.,
  {Barclay} T.~S.,  {Fanelli} M.~N.,  {Girouard} F.~R.,  {Kolodziejczak} J.~J.,
   {McCauliff} S.~D.,  {Morris} R.~L.,    {Twicken} J.~D.,  2012, PASP, 124,
  1000

\bibitem[\protect\citeauthoryear{{Storn} \& {Price}}{{Storn} \&
  {Price}}{1997}]{sp97}
{Storn} R.,  {Price} K.,  1997, Journal of Global Optimization, 11, 341

\bibitem[\protect\citeauthoryear{{Stumpe}, {Smith}, {Catanzarite}, {Van Cleve},
  {Jenkins}, {Twicken} \& {Girouard}}{{Stumpe} et~al.}{2014}]{stu+14}
{Stumpe} M.~C.,  {Smith} J.~C.,  {Catanzarite} J.~H.,  {Van Cleve} J.~E.,
  {Jenkins} J.~M.,  {Twicken} J.~D.,    {Girouard} F.~R.,  2014, PASP, 126, 100

\bibitem[\protect\citeauthoryear{{Stumpe}, {Smith}, {Van Cleve}, {Twicken},
  {Barclay}, {Fanelli}, {Girouard}, {Jenkins}, {Kolodziejczak}, {McCauliff} \&
  {Morris}}{{Stumpe} et~al.}{2012}]{stu+12}
{Stumpe} M.~C.,  {Smith} J.~C.,  {Van Cleve} J.~E.,  {Twicken} J.~D.,
  {Barclay} T.~S.,  {Fanelli} M.~N.,  {Girouard} F.~R.,  {Jenkins} J.~M.,
  {Kolodziejczak} J.~J.,  {McCauliff} S.~D.,    {Morris} R.~L.,  2012, PASP,
  124, 985

\bibitem[\protect\citeauthoryear{{Tamuz}, {Mazeh} \& {Zucker}}{{Tamuz}
  et~al.}{2005}]{tam+05}
{Tamuz} O.,  {Mazeh} T.,    {Zucker} S.,  2005, MNRAS, 356, 1466

\bibitem[\protect\citeauthoryear{{Tenenbaum}, {Christiansen}, {Jenkins},
  {Rowe}, {Seader}, {Caldwell}, {Clarke}, {Li}, {Quintana}, {Smith}, {Stumpe},
  {Thompson}, {Twicken}, {Van Cleve}, {Borucki}, {Cote} \& {et
  al.}}{{Tenenbaum} et~al.}{2012}]{ten+12}
{Tenenbaum} P.,  {Christiansen} J.~L.,  {Jenkins} J.~M.,  {Rowe} J.~F.,
  {Seader} S.,  {Caldwell} D.~A.,  {Clarke} B.~D.,  {Li} J.,  {Quintana} E.~V.,
   {Smith} J.~C.,  {Stumpe} M.~C.,  {Thompson} S.~E.,  {Twicken} J.~D.,  {Van
  Cleve} J.,  {Borucki} W.~J.,  {Cote} M.~T.,    {et al.} 2012, ApJS, 199, 24

\bibitem[\protect\citeauthoryear{{Van Cleve}, {Howell}, {Smith}, {Clarke},
  {Thompson}, {Bryson}, {Lund}, {Handberg} \& {Chaplin}}{{Van Cleve}
  et~al.}{2015}]{van+15}
{Van Cleve} J.~E.,  {Howell} S.~B.,  {Smith} J.~C.,  {Clarke} B.~D.,
  {Thompson} S.~E.,  {Bryson} S.~T.,  {Lund} M.~N.,  {Handberg} R.,
  {Chaplin} W.~J.,  2015, PASP, in press, arXiv:1512.06162

\bibitem[\protect\citeauthoryear{{Vanderburg}}{{Vanderburg}}{2014}]{van14}
{Vanderburg} A.,  2014, ArXiv e-prints

\bibitem[\protect\citeauthoryear{{Vanderburg} \& {Johnson}}{{Vanderburg} \&
  {Johnson}}{2014}]{vj14}
{Vanderburg} A.,  {Johnson} J.~A.,  2014, PASP, 126, 948

\end{thebibliography}
\bibliographystyle{mn2e}

\bsp

\label{lastpage}

\end{document}